\documentclass[twocolumn,apl,reprint,showpacs,preprintnumbers,amsmath,amssymb,superscriptaddress]{revtex4-1}
\usepackage{graphicx}
\usepackage{dcolumn}
\usepackage{bm}

\usepackage{amsmath}
\usepackage{subfigure}
\usepackage{hyperref}
\usepackage[normalem]{ulem}
\hypersetup{
    colorlinks,
    citecolor=blue,
    filecolor=black,
    linkcolor=blue,
    urlcolor=blue
}
\usepackage{epstopdf}
 \usepackage{relsize}

\newcommand*{\rom}[1]{\expandafter\@slowromancap\romannumeral #1@}
\begin{document}


\title{Optimized Peltier cooling via an array of quantum dots with stair-like ground-state energy configuration}

\author{Aniket Singha}
\email{aniket@ece.iitkgp.ac.in}
\affiliation{%
Department of Electronics and Electrical Communication Engineering,\\
Indian Institute of Technology Kharagpur,  Kharagpur-721302, India\\
}%


%



\date{\today}

\begin{abstract}
With the advancement in fabrication  and scaling technology, the rising temperature in nano devices has attracted special attention towards thermoelectric or Peltier cooling. In this paper, I propose optimum Peltier cooling by employing an array of connected quantum dots with stair-like ground-state eigen energy configuration. The difference in ground state eigen energy  between two adjacent quantum dots in the stair-like configuration is chosen to be identical with the optical phonon energy for efficient absorption of lattice heat. I show that in the proposed configuration, for a given  optical phonon energy, one can optimize the cooling power by tuning the number of stages in the array of quantum dots. A further analysis    demonstrates that the   maximum cooling power at a given potential bias under optimal conditions doesnot depend strongly on the optical phonon energy or the number of stages at which the maximum cooling power is achieved, provided that the optical phonon energy is  less than $kT$.  The proposed concept can also be applied to $2-D$ or bulk resonant tunnel and superlattice structures with stair-like resonant energy configuration.
\end{abstract}
\maketitle
\section{Introduction}
With the recent advancement in fabrication and scaling technology, the rising dissipated heat density and temperature in nano devices have drawn special attention towards Peltier cooling. Peltier cooling, the reverse phenomenon of thermoelectric generation \cite{goldsmid,extra1,extra2,jordan1,jordan2, response1,response2,response3,response4,sofo},  is achieved by an energy selective flow of electronic current via energy filtering. This results in a disturbance of quasi-equilibrium within a given energy range initiating abrorption of lattice heat in an attempt to establish equilibrium among the electronic population \cite{snyder_thompson,cooling_ref2,cooling_ref5,cooling_ref6,cooling_ref9,cooling_ref10,cooling_ref11,whitney2,whitney}.  Peltier coolers are mainly attractive because in addition to being light weight and cost efficient, they are also less prone to failure. Resonant tunneling and superlattice based Peltier coolers have been theoretically proposed as well as experimentally realized \cite{two,snyder_thompson,cooling_ref1,cooling_ref2,cooling_ref3,cooling_ref4,cooling_ref5,cooling_ref6,cooling_ref7,cooling_ref8,cooling_ref9,cooling_ref10,cooling_ref11}.      However, the principal drawback of such coolers is inefficient absorption of heat energy due to a mismatch between the optical phonon energy and  eigen energy   (resonant energy) of adjacent quantum dots (resonant tunnel structures) .  In this paper, I propose an array of quantum dots with stair-like ground-state eigen  energy configuration embedded in a nanowire-like structure  to optimize cooling performance. \\ \indent The motivation of this proposal stems from the fact that the absorption or emission rate of phonons in  a Peltier cooler peak when the difference in energy levels between two adjacent quantum dots is identical to the phonon energy (this can be shown starting from the electron-phonon interaction Hamiltonian) \cite{lundstorm,dattabook,Datta_Green,fetter,sakurai,griffith}.  However, when the difference in eigen energy is not identical to the optical phonon energy, the cooling performance deteriorates from the peak value. The aspect of thermoelectric generation and cooling using an array of two quantum dots with different ground state energy has already been discussed in literature \cite{two}. However, the optimization of cooling power based on the  phonon energy and temperature by tuning (increasing) the number of stages of the quantum dot array is, indeed, new and deserves special attention.\\
\indent In this paper, I propose that one can optimize the cooling performance by tuning the number of stages of a quantum dot array while keeping the ground state eigen-energy difference between the adjacent quantum dots identical to the optical phonon energy and the temperature. I show that in such an arrangement, the magnitude of optical phonon energy has minimal impact on the optimized cooling performance provided that the optical phonon energy is lower than $kT$. The optimum number of quantum dots in the array would, however, depend on the optical phonon energy.   Although, in this paper, I perform an investigation of the cooling performance harnessed from an array of quantum dots, the proposed concept can be applied in $2-D$ and bulk resonant tunnel or superlattice structures with stair-like resonant energy configuration. \\
\begin{figure}[!htb]
\centering
\hspace{-.5cm} \includegraphics[scale=.12]{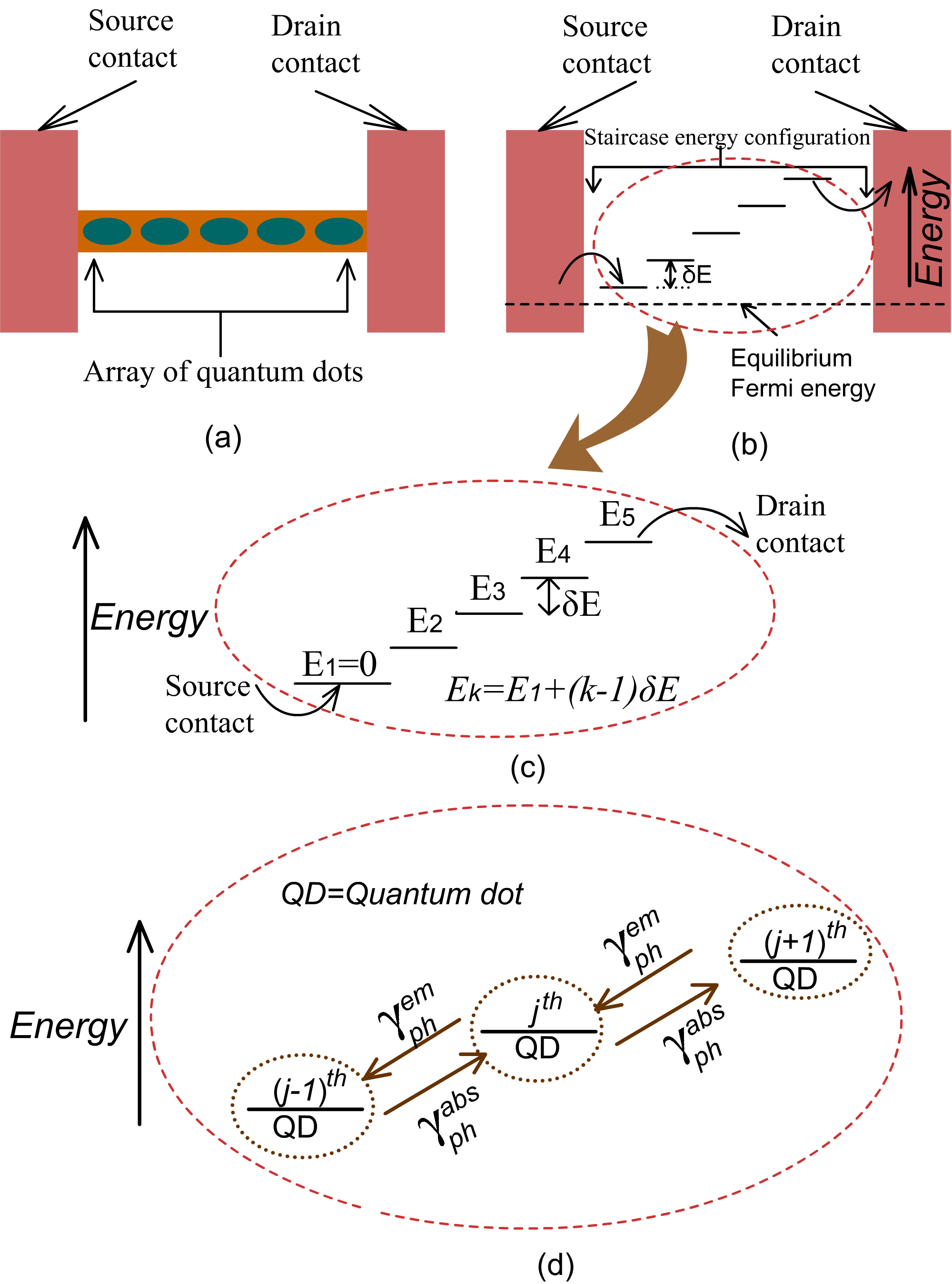}
\caption{Device schematics for an array of quantum dots with stair-like ground-state eigen energy configuration. (a) Schematic of the proposed arrangement depicting a nanowire embedded with an array of quantum dots. (b) Schematic of stair-like ground-state energy configuration in the array of quantum dots. (c) Magnified schematic of the stair-like ground-state eigen energy configuration in the array of quantum dots. The ground state eigen energy $E_1$ of the first quantum dot is assumed to be $0$ (d) Schematic of the transport formulation used in this paper.}
\label{fig:schematic}
\end{figure}
\begin{figure}
\centering
\subfigure[]{ \includegraphics[scale=.3]{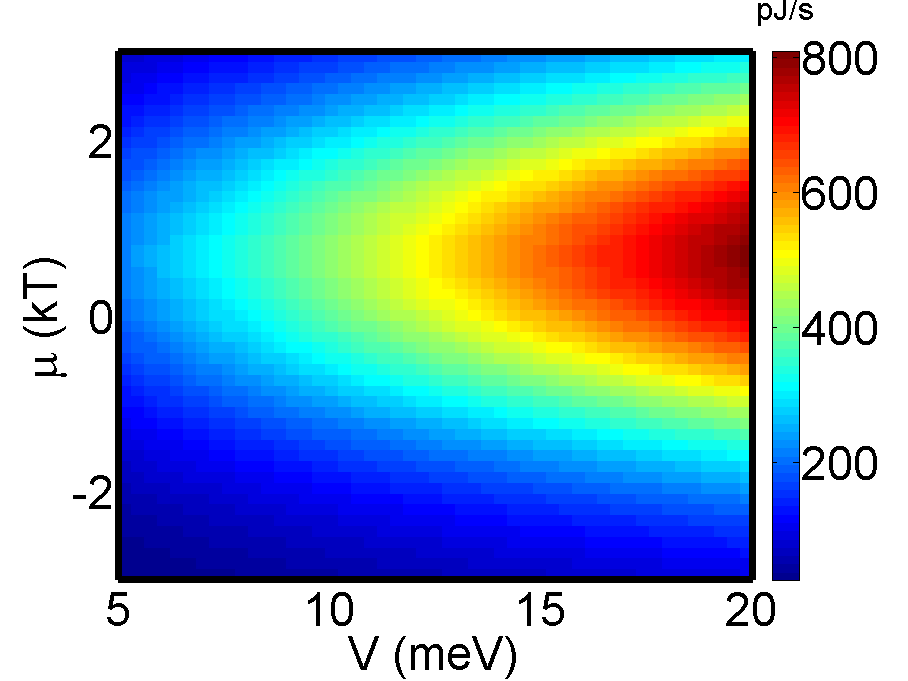}}
\subfigure[]{ \includegraphics[scale=.3]{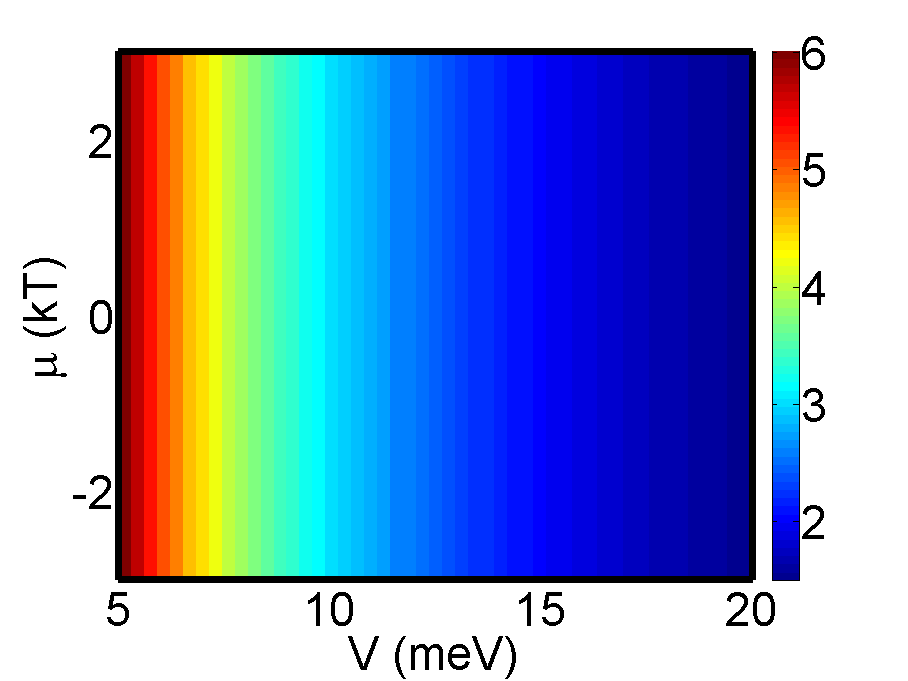}}
\caption{Analysis of  cooling performance in the proposed arrangement with an array of two quantum dots with $\delta E=\Delta E=30meV$. Colour plot showing  the variation of (a) cooling power with the position of the equilibrium Fermi energy and the applied bias voltage (b) the coefficient of performance (COP)  with the position of the equilibrium Fermi energy and the applied bias voltage.}
\label{fig:sec_fig}
\end{figure}
\section{Proposed Device Structure}
\indent Fig. \ref{fig:schematic} demonstrates the  schematic diagram of the proposed 	Peltier cooler employing an array of $n$ quantum dots. The  first and the last quantum dots are coupled to the left (source) and right (drain) contact respectively. The rate of inflow and outflow of electrons from the contacts to the dots is assumed to be $\gamma_c$. The  flow of electrons between two adjacent quantum dots consists of two different mechanisms: ($i$) Elastic tunneling between the dots without exchange of energy between the electrons and the lattice  ($ii$) Inelastic or phonon assisted tunneling between two dots via absorption  (emission) of energy from the lattice. A staircase ground state energy configuration in an array of quantum dots aids in reducing the elastic tunneling rate compared to inelastic tunneling rate thereby improving the coefficient of performance of the Peltier cooler. \\
\begin{figure*}[!htb]
\centering
\hspace{-2cm} \includegraphics[scale=.45]{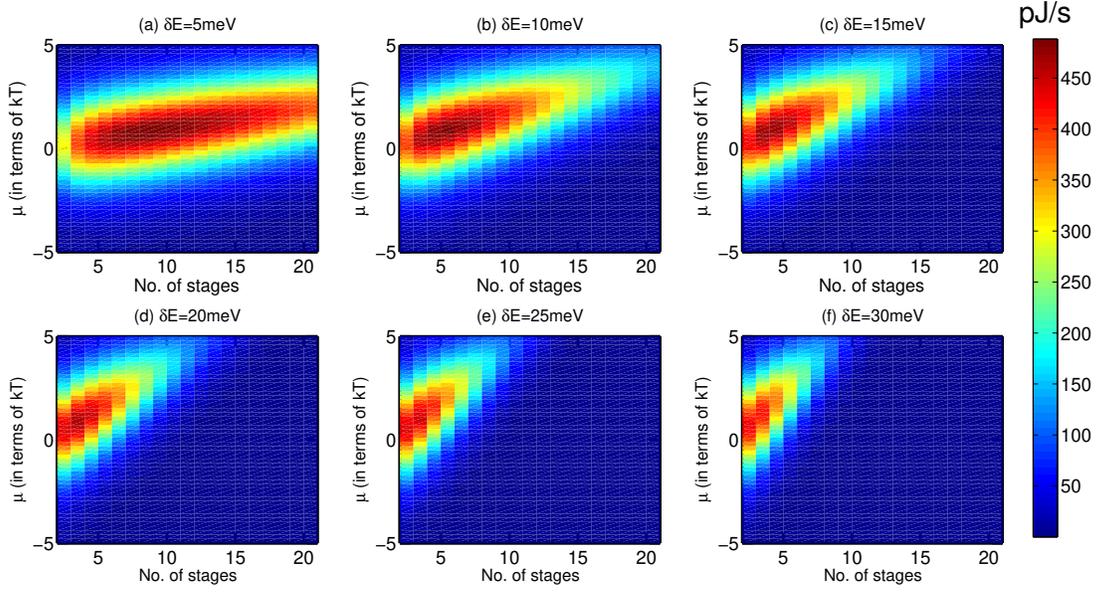}
\caption{Variation of the cooling power with the number of stages of the array of quantum dots for different optical phonon energy ($\hbar \omega=\delta E$) at $V=10mV$ and $T=300K$. The difference in ground-state energy of adjacent quantum dots is assumed to be equal to the optical phonon energy ($\delta E=\hbar \omega$). Plots shown for (a) $\delta E=5meV$ (b) $\delta E=10meV$ (c) $\delta E=15meV$ (d) $\delta E=20meV$ (e) $\delta E=25meV$ (f) $\delta E=30meV$}
\label{fig:third_fig}
\end{figure*}
\begin{figure}
\centering
\subfigure[]{ \includegraphics[scale=.25]{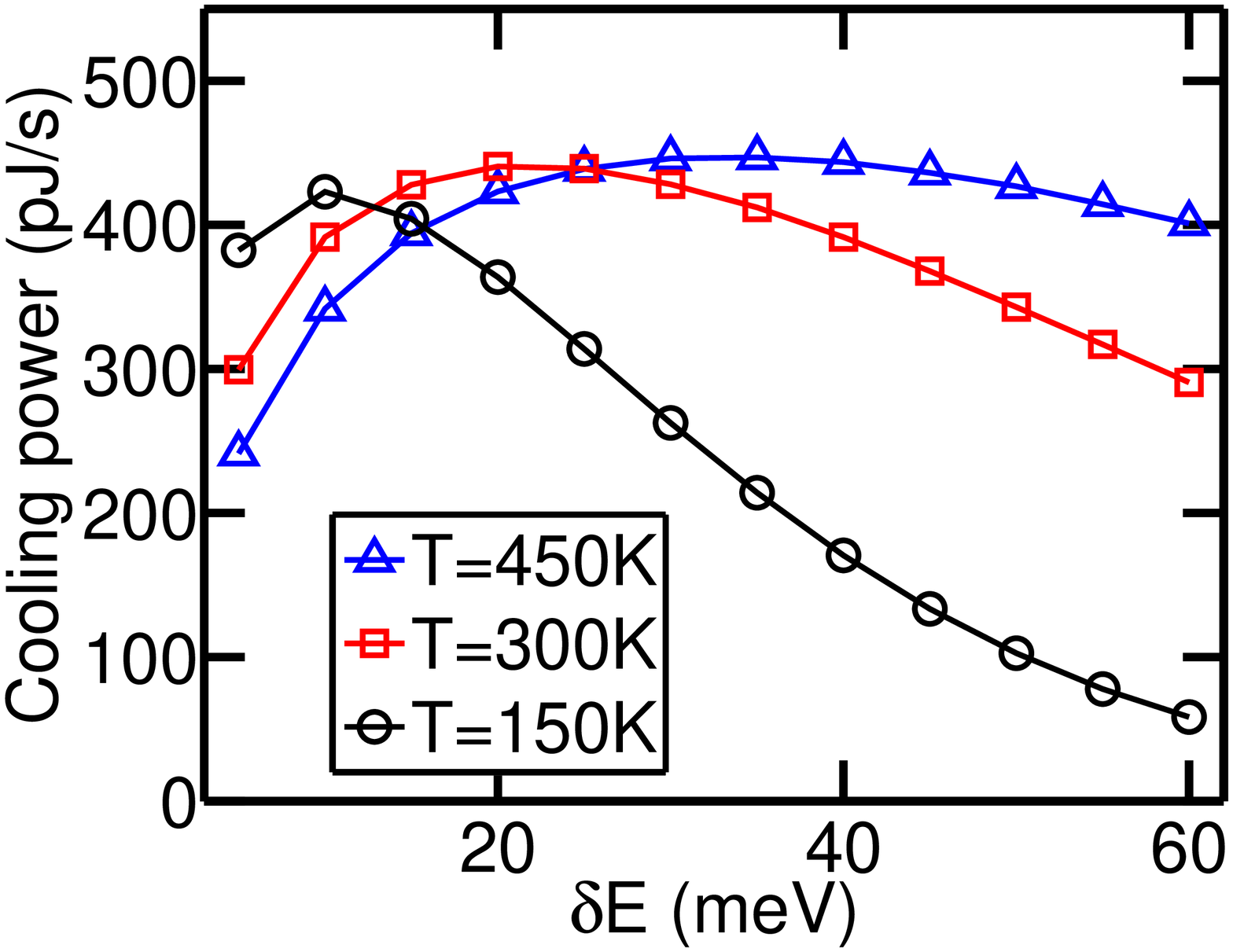}}
\subfigure[]{ \includegraphics[scale=.25]{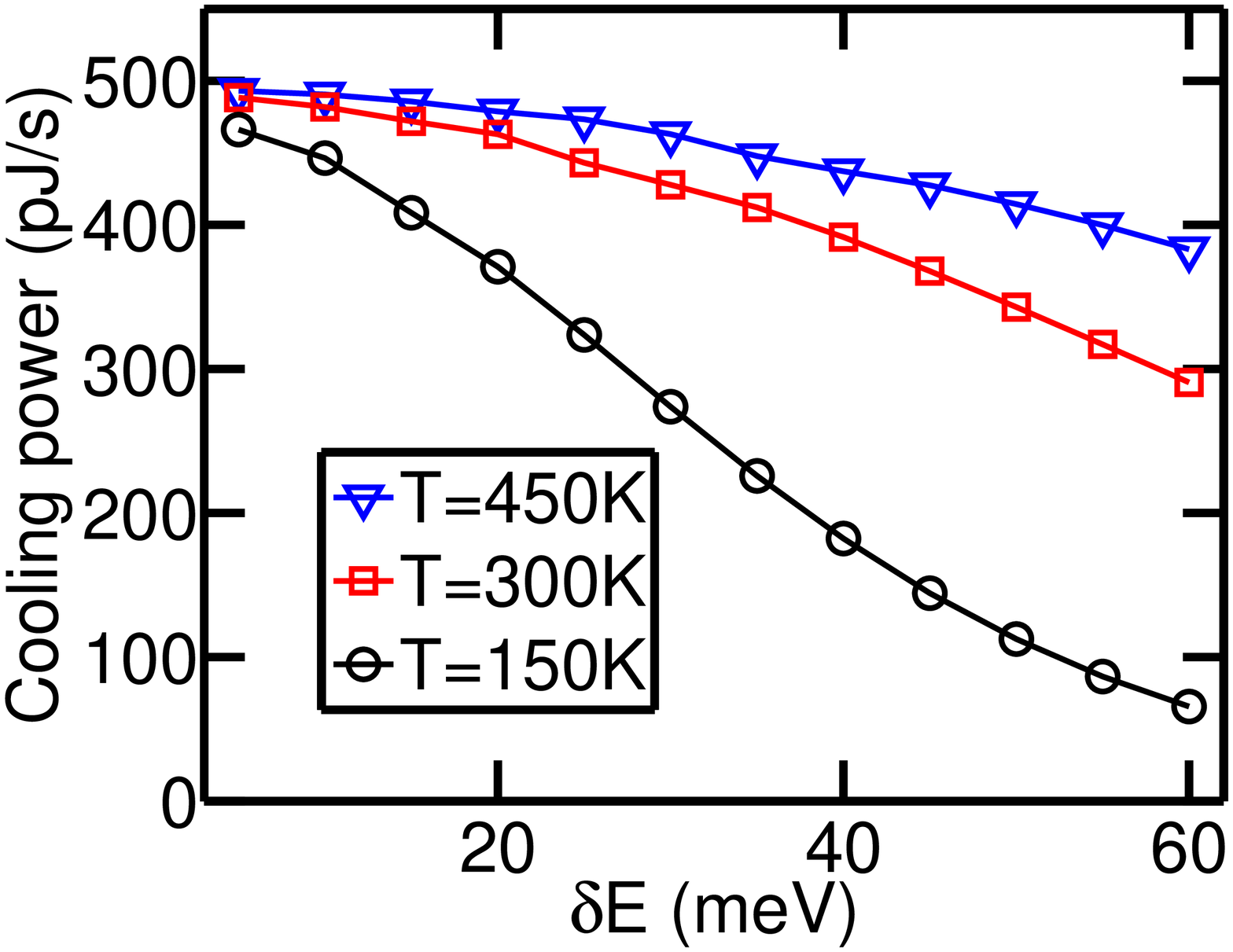}}
\subfigure[]{ \includegraphics[scale=.25]{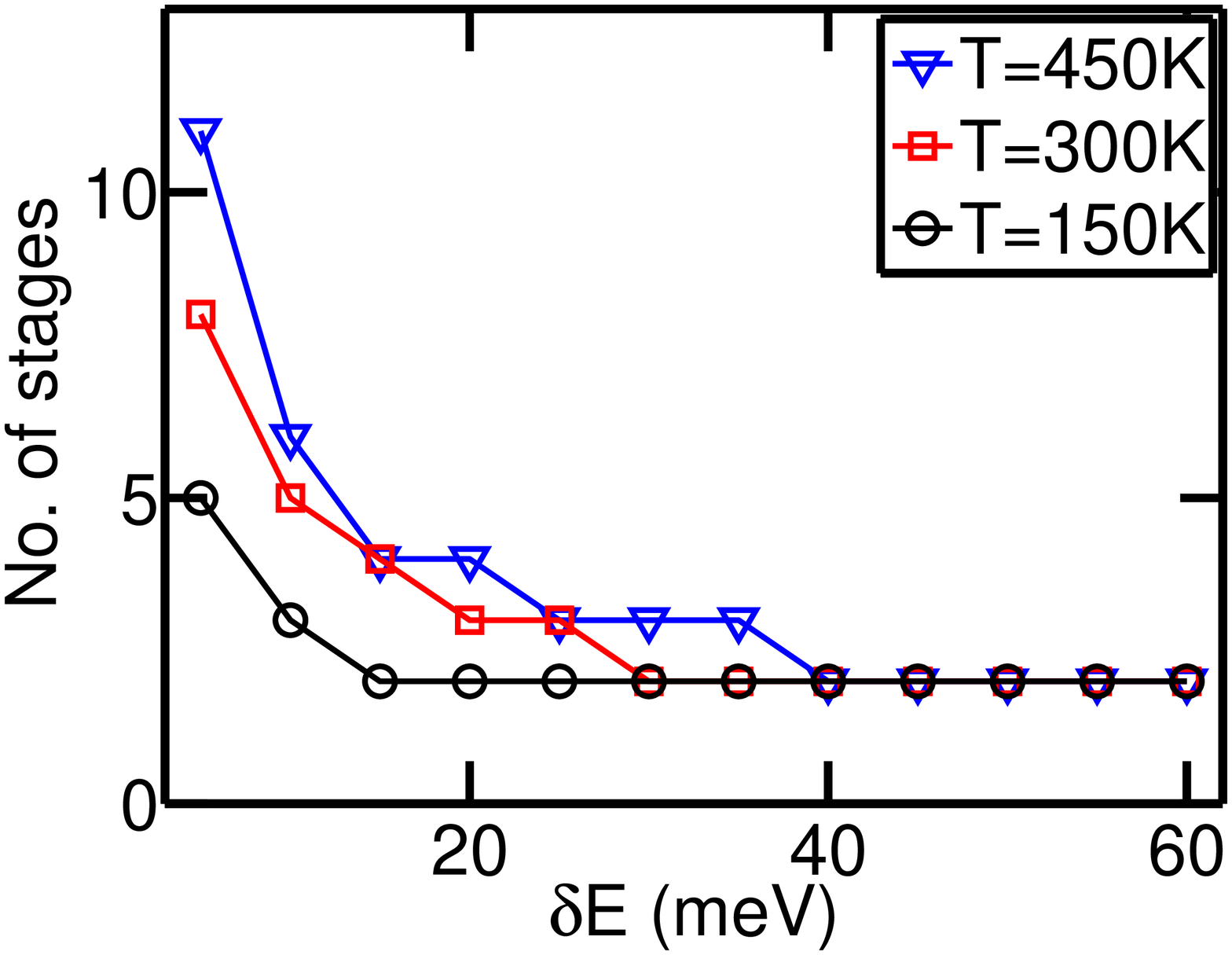}}
\caption{Analysis of optimal cooling performance in the proposed arrangement at $V=10mV$. (a) The maximum cooling power as a function of the optical phonon energy $\delta E (\hbar \omega)$ using an array of two quantum dots. (b) The maximum cooling power as a function of the optical phonon energy $\delta E (\hbar \omega)$ using the optimized number of quantum dots in the array (c) The optimum  number of quantum dots which maximizes the cooling power as a function of the optical phonon energy $\delta E (\hbar \omega)$}
\label{fig:fourth_fig}
\end{figure}
\section{Analysis methodology and transport formulation}
\indent  There are two major pathways to improve the  performance of Peltier coolers (i) decreasing the lattice thermal conductance (ii) enhancing the cooling power via suitable energy filtering techniques. Both of these techniques are widely employed to enhance the performance of thermoelectric generators  \cite{goldsmid,extra1,extra2,jordan1,jordan2, response1,response2,response3,response4} and coolers \cite{two,snyder_thompson,cooling_ref1,cooling_ref2,cooling_ref3,cooling_ref4,cooling_ref5,cooling_ref6,cooling_ref7,cooling_ref8,cooling_ref9,cooling_ref10,cooling_ref11}.Approaches towards the direction of nanostructuring as well as heterostructuring has so far proven successful in the suppressing phonon mediated lattice thermal conductivity via scattering and confinement of long wavelength phonons \cite{phonon1,phonon2,phonon3,phonon4,phonon5,phonon6,phonon7,superlattice1,superlattice2,nanoflake_heat,nanowire_heat1,nanowire_heat2}. Hence, in this paper, we focus on optimizing the cooling power without the consideration for lattice thermal conductance.\\
\indent Without loss of generality, I will assume that the rate of inelastic tunneling between adjacent quantum dots via energy absorption (emission) is denoted by the parameter $\gamma_{ph}^{abs (em)}$, while elastic tunneling between adjacent quantum dots is assumed to be negligible due to the stair-case energy configuration. The rate of electronic transfer between the contacts and the leftmost (rightmost) quantum dot, on the other hand, is denoted by $\gamma_c$.  Assuming equilibrium among phonon population, the rate of inelastic tunneling ($\gamma_{ph}^{abs (em)}$) between adjacent quantum dots is related to the optical phonon number $\left(N_{ph}=\left[{exp(\frac{\delta E}{kT})-1}\right]^{-1}\right)$ as:
\begin{equation}
\gamma_{ph}^{abs (em)}=t_{ph} \times \left[ N_{ph}+\frac{1}{2} -(+)\frac{1}{2}\right],
\end{equation} 
where $t_{ph}$ is related to the coupling between adjacent quantum dots due to optical phonon scattering. A schematic diagram depicting the transport formulation is shown in Fig. \ref{fig:schematic}~(d). The temporal rate of change of probability of electronic occupancy in a particular  quantum dot is dependent on the probability of occupancy of the adjacent quantum dots and is   described via master equation. The first and the last quantum dots are connected directly to the contacts. In these cases, the rate of change of occupancy probability of the first quantum dot ($P_1$)  is  given by:
\begin{multline}
\frac{dP_1}{dt}= 
    \gamma_c f_L (1-P_{1})+\gamma_{ph}^{em} P_{2} (1-P_{1}) \\ -\gamma_{ph}^{abs} P_{1} (1-P_{2})-\gamma_c P_{1} (1-f_L), 
    \label{eq:P_1}
\end{multline}
while that for the $n^{th}$ quantum dot ($P_n$) is given by:
\begin{multline}
    \frac{dP_n}{dt}= 
    \gamma_{ph}^{abs} P_{n-1} (1-P_{n})+\gamma_c f_R (1-P_{n}) \\ -\gamma_c P_{n} (1-f_R)-\gamma_{ph}^{em} P_{n} (1-P_{n-1}). 
\label{eq:P_n}
\end{multline}
The probability of electronic occupancy for the $j^{th}$ quantum dot ($1<j<n$) is given by:
\begin{multline}
\frac{dP_j}{dt}= 
    \gamma_{ph}^{abs} P_{j-1} (1-P_{j})+\gamma_{ph}^{em} P_{j+1} (1-P_{j})\\ -\gamma_{ph}^{abs} P_{j} (1-P_{j+1})-\gamma_{ph}^{em} P_{j} (1-P_{j-1}). 
\label{eq:P_j}
\end{multline}
where $f_L=\left[{1+exp\left(-\frac{\mu+V/2}{kT}\right)}\right]^{-1}$ is the electronic occupancy at the energy corresponding to the ground-state of the $1^{st}$ quantum dot and $f_R=\left[{1+exp\left(\frac{\Delta E-(\mu-V/2)}{kT}\right)}\right]^{-1}$ is the electronic occupancy at the energy corresponding to the ground-state of the $n^{st}$ quantum dot, $\mu$ being the position of the equilibrium Fermi energy and $\Delta E=E_n=(n-1)\delta E$ defines  the ground-state eigen energy of the $n^{th}$ quantum dot ($E_1=0$). In steady state, $\frac{dP_j}{dt}=0~~(1\leq j \leq n)$.
The interdependence among the  set of Eqs. \eqref{eq:P_1}-\eqref{eq:P_j} is solved self-consistently using Newton-Raphson  method assuming $\frac{dP_j}{dt}=0~~(1\leq j \leq n)$.  The charge current at the left (source) and right (drain) contacts are given by:
\begin{eqnarray}
I_L=2q\left[\gamma_c(1-P_1)f_L-\gamma_c P_1 (1-f_L)\right] \nonumber \\
I_R=2q\left[\gamma_c(1-P_n)f_R-\gamma_c P_n (1-f_R)\right],
\label{eq:curr}
\end{eqnarray}
where $V$ is the applied voltage bias across the array of quantum dots. The multiplicative factor of $2$ takes  the spin degeneracy into account.
The heat  extracted at the $j^{th}$ quantum dot may be given by the equation
\begin{equation}
I^Q_j=2\delta E \times\{ \gamma_{ph}^{abs} P_j (1-P_{j+1})-\gamma_{ph}^{em} (1-P_j) P_{j+1} \}
\label{eq:heat}
\end{equation}
The total heat current extracted from the system is the sum of the heat extracted from individual quantum dots.
\begin{equation}
I^Q_{total}=\sum_j I^Q_j
\end{equation}
The efficiency of the Peltier cooler is related to the coefficient of performance (COP) defined as:
\begin{equation}
COP=\frac{I^Q_{total}}{V \times I_{L (R)}}.
\end{equation}
 A list of parameters used for simulation is given in TABLE.~I. 
\begin{table}[!htb]
\caption{Parameters used for simulation.}
\begin{center}
    \begin{tabular}{ | p{2cm} | p{6cm} |}
    \hline
    \textbf{Parameters} & \textbf{Values}  \\ \hline \hline
    $T~~(kT)$  & $300K~~(25.85~meV)$  \\ \hline
    $\gamma_c$  & $5\times 10^{-3} {q}/{\hbar}~sec^{-1}$  \\ \hline
    $t_{ph}$ & $5\times 10^{-4} {q}/{\hbar}~sec^{-1}$  \\ \hline
        $E_1$ & $0$  \\ \hline

    \end{tabular}
       \vspace{1ex}
\end{center}
\end{table} \\

\section{Results}
\indent Fig.~\ref{fig:sec_fig} demonstrates the cooling performance of an array of two quantum dots with $\Delta E=\delta E=\hbar \omega=30meV$. For a given voltage bias, the cooling power is maximum around $\mu=kT$ and increases with the applied voltage bias. The coefficient of performance on the other hand  is a decreasing function of the applied voltage for a given position of the equilibrium Fermi energy. Fig \ref{fig:third_fig} demonstrates the cooling power histogram over a range of the equilibrium Fermi energy and the number of stages of the array of quantum dot for several values of the optical phonon energy ($\hbar \omega=\delta E$) at $V=10mV$ and $T=300K$. The number of stages in the array of quantum dots for optimal cooling power decreases with the increase in optical phonon energy. However, the maximum cooling power is not strongly dependent on the optical phonon energy.  Fig. \ref{fig:fourth_fig} demonstrates the optimal characteristics of the proposed cooler at $V=10mV$ for various operating temperatures. In particular, Fig. \ref{fig:fourth_fig}(a) demonstrates the maximum cooling power as a function of the  optical phonon energy with an array of two quantum dots.  Fig. \ref{fig:fourth_fig}(b), on the other hand,  displays the maximum cooling power of the proposed cooler with variation in optical phonon energy when the number of quantum dots in the array is optimized. Clearly, we get an improved cooling power in the latter case for the regime $\hslash \omega<kT$. Fig. \ref{fig:fourth_fig}(c) shows the number of stages of the proposed cooler at which maximum cooling power is achieved as a function of optical phonon energy ($\delta E=\hbar \omega$). It is clear from Fig.~\ref{fig:fourth_fig} that the maximum cooling power using an array of two quantum dots (shown in Fig. \ref{fig:fourth_fig}~a) is identical to the optimized cooling power (Fig. \ref{fig:fourth_fig}~b) when the optical phonon energy is comparable or high compared to $kT$. On the other hand, when the optical phonon energy is low, the cooling power using an optimized number of quantum dots deviates significantly compared to the the achieved optimal cooling power.  Although the number of stages for optimal cooling in the array of quantum dots increases with decrease in optical phonon energy, such an increase doesnot suppress the maximum cooling power. This is because  a decrease in the optical phonon energy also increases the average  phonon number ($N_{ph}$) which causes an increase in the average inelastic tunneling between adjacent quantum dots. This phenomenon compensates for the increase in the number of stages with decrease in optical phonon energy when $\delta E <kT$.   On the other hand, for $\delta E \geq  kT$ the cooling power varies sharply with the optical phonon energy due to an exponential decrease in the phonon number $N_{ph}$ with energy. As the optical phonon energy ($\hbar \omega$) increases towards $kT$, the optimal number of stages in the array of quantum dots reaches the limiting value of $2$. It should be noted that the minimum number of quantum dots required in the proposed arrangement is $2$ \cite{two}.  \\
\section{Conclusion}
\indent To conclude, in this paper, I have proposed a new arrangement for Peltier cooling with the help of an array of quantum dots with stair-like ground state  energy configuration. It was shown that the cooling power of such a cooler improves compared to the traditional case ($n=2$) for $\hslash \omega<kT$. An improvement in cooling power also facilitates a larger drop in temperature. It was also shown that the performance of such a Peltier cooler, in terms of the maximum cooling power, is almost independent of the optical phonon energy provided that $\hslash \omega < kT$ and the number of stages in the array  is optimized. Our considerations and results presented here are based on the assumption that  the optical phonon energy is greater than the eigen energy broadening of the quantum dots, such that the effect of elastic tunneling can be suppressed. Elastic tunneling degrades the peak performance of such a refrigerator. Particularly, it would  result in the degradation of the maximum cooling power and the coefficient of performance. Investigation of the effect of elastic tunneling in the limit of low optical phonon energy is left for future work. In addition, an investigation of the temperature drop in such a refrigerator under various conditions also constitute an interesting research problem. Although the proposed arrangement involves employment of an array of quantum dots embedded in a nanowire like structure, the concept presented is   general and can be  applied  for optimal cooling in resonant tunnel  or super-lattice   structures with stair-like resonant energy configuration. The proposed Peltier cooler would lead enhancement in performance of experimentally fabricated coolers.
\vfill

\bibliography{apssamp}

\providecommand{\noopsort}[1]{}\providecommand{\singleletter}[1]{#1}%
\begin{thebibliography}{43}%
\makeatletter
\providecommand \@ifxundefined [1]{%
 \@ifx{#1\undefined}
}%
\providecommand \@ifnum [1]{%
 \ifnum #1\expandafter \@firstoftwo
 \else \expandafter \@secondoftwo
 \fi
}%
\providecommand \@ifx [1]{%
 \ifx #1\expandafter \@firstoftwo
 \else \expandafter \@secondoftwo
 \fi
}%
\providecommand \natexlab [1]{#1}%
\providecommand \enquote  [1]{``#1''}%
\providecommand \bibnamefont  [1]{#1}%
\providecommand \bibfnamefont [1]{#1}%
\providecommand \citenamefont [1]{#1}%
\providecommand \href@noop [0]{\@secondoftwo}%
\providecommand \href [0]{\begingroup \@sanitize@url \@href}%
\providecommand \@href[1]{\@@startlink{#1}\@@href}%
\providecommand \@@href[1]{\endgroup#1\@@endlink}%
\providecommand \@sanitize@url [0]{\catcode `\\12\catcode `\$12\catcode
  `\&12\catcode `\#12\catcode `\^12\catcode `\_12\catcode `\%12\relax}%
\providecommand \@@startlink[1]{}%
\providecommand \@@endlink[0]{}%
\providecommand \url  [0]{\begingroup\@sanitize@url \@url }%
\providecommand \@url [1]{\endgroup\@href {#1}{\urlprefix }}%
\providecommand \urlprefix  [0]{URL }%
\providecommand \Eprint [0]{\href }%
\providecommand \doibase [0]{http://dx.doi.org/}%
\providecommand \selectlanguage [0]{\@gobble}%
\providecommand \bibinfo  [0]{\@secondoftwo}%
\providecommand \bibfield  [0]{\@secondoftwo}%
\providecommand \translation [1]{[#1]}%
\providecommand \BibitemOpen [0]{}%
\providecommand \bibitemStop [0]{}%
\providecommand \bibitemNoStop [0]{.\EOS\space}%
\providecommand \EOS [0]{\spacefactor3000\relax}%
\providecommand \BibitemShut  [1]{\csname bibitem#1\endcsname}%
\let\auto@bib@innerbib\@empty
\bibitem [{\citenamefont {Goldsmid}(2009)}]{goldsmid}%
  \BibitemOpen
  \bibfield  {author} {\bibinfo {author} {\bibnamefont {Goldsmid}},\
  }\href@noop {} {\emph {\bibinfo {title} {{Introduction to Thermoelectricity
  }}}}\ (\bibinfo  {publisher} {Springer},\ \bibinfo {year} {2009})\BibitemShut
  {NoStop}%
\bibitem [{\citenamefont {DiSalvo}(1999)}]{extra1}%
  \BibitemOpen
  \bibfield  {author} {\bibinfo {author} {\bibfnamefont {F.~J.}\ \bibnamefont
  {DiSalvo}},\ }\href {\doibase 10.1126/science.285.5428.703} {\bibfield
  {journal} {\bibinfo  {journal} {Science}\ }\textbf {\bibinfo {volume}
  {285}},\ \bibinfo {pages} {703} (\bibinfo {year} {1999})}\BibitemShut
  {NoStop}%
\bibitem [{\citenamefont {Shakouri}(2011)}]{extra2}%
  \BibitemOpen
  \bibfield  {author} {\bibinfo {author} {\bibfnamefont {A.}~\bibnamefont
  {Shakouri}},\ }\href {\doibase 10.1146/annurev-matsci-062910-100445}
  {\bibfield  {journal} {\bibinfo  {journal} {Annual Review of Materials
  Research}\ }\textbf {\bibinfo {volume} {41}},\ \bibinfo {pages} {399}
  (\bibinfo {year} {2011})}\BibitemShut {NoStop}%
\bibitem [{\citenamefont {Jordan}\ \emph {et~al.}(2013)\citenamefont {Jordan},
  \citenamefont {Sothmann}, \citenamefont {S\'anchez},\ and\ \citenamefont
  {B\"uttiker}}]{jordan1}%
  \BibitemOpen
  \bibfield  {author} {\bibinfo {author} {\bibfnamefont {A.~N.}\ \bibnamefont
  {Jordan}}, \bibinfo {author} {\bibfnamefont {B.}~\bibnamefont {Sothmann}},
  \bibinfo {author} {\bibfnamefont {R.}~\bibnamefont {S\'anchez}}, \ and\
  \bibinfo {author} {\bibfnamefont {M.}~\bibnamefont {B\"uttiker}},\ }\href
  {\doibase 10.1103/PhysRevB.87.075312} {\bibfield  {journal} {\bibinfo
  {journal} {Phys. Rev. B}\ }\textbf {\bibinfo {volume} {87}},\ \bibinfo
  {pages} {075312} (\bibinfo {year} {2013})}\BibitemShut {NoStop}%
\bibitem [{\citenamefont {Choi}\ and\ \citenamefont {Jordan}(2015)}]{jordan2}%
  \BibitemOpen
  \bibfield  {author} {\bibinfo {author} {\bibfnamefont {Y.}~\bibnamefont
  {Choi}}\ and\ \bibinfo {author} {\bibfnamefont {A.~N.}\ \bibnamefont
  {Jordan}},\ }\href {\doibase http://dx.doi.org/10.1016/j.physe.2015.08.002}
  {\bibfield  {journal} {\bibinfo  {journal} {Physica E: Low-dimensional
  Systems and Nanostructures}\ ,\ } (\bibinfo {year} {2015})}\BibitemShut
  {NoStop}%
\bibitem [{\citenamefont {Giazotto}\ \emph
  {et~al.}(2006{\natexlab{a}})\citenamefont {Giazotto}, \citenamefont
  {Heikkil\"a}, \citenamefont {Luukanen}, \citenamefont {Savin},\ and\
  \citenamefont {Pekola}}]{response1}%
  \BibitemOpen
  \bibfield  {author} {\bibinfo {author} {\bibfnamefont {F.}~\bibnamefont
  {Giazotto}}, \bibinfo {author} {\bibfnamefont {T.~T.}\ \bibnamefont
  {Heikkil\"a}}, \bibinfo {author} {\bibfnamefont {A.}~\bibnamefont
  {Luukanen}}, \bibinfo {author} {\bibfnamefont {A.~M.}\ \bibnamefont {Savin}},
  \ and\ \bibinfo {author} {\bibfnamefont {J.~P.}\ \bibnamefont {Pekola}},\
  }\href {\doibase 10.1103/RevModPhys.78.217} {\bibfield  {journal} {\bibinfo
  {journal} {Rev. Mod. Phys.}\ }\textbf {\bibinfo {volume} {78}},\ \bibinfo
  {pages} {217} (\bibinfo {year} {2006}{\natexlab{a}})}\BibitemShut {NoStop}%
\bibitem [{\citenamefont {Muhonen}\ \emph {et~al.}(2012)\citenamefont
  {Muhonen}, \citenamefont {Meschke},\ and\ \citenamefont
  {Pekola}}]{response2}%
  \BibitemOpen
  \bibfield  {author} {\bibinfo {author} {\bibfnamefont {J.~T.}\ \bibnamefont
  {Muhonen}}, \bibinfo {author} {\bibfnamefont {M.}~\bibnamefont {Meschke}}, \
  and\ \bibinfo {author} {\bibfnamefont {J.~P.}\ \bibnamefont {Pekola}},\
  }\href {http://stacks.iop.org/0034-4885/75/i=4/a=046501} {\bibfield
  {journal} {\bibinfo  {journal} {Reports on Progress in Physics}\ }\textbf
  {\bibinfo {volume} {75}},\ \bibinfo {pages} {046501} (\bibinfo {year}
  {2012})}\BibitemShut {NoStop}%
\bibitem [{\citenamefont {Paulsson}\ and\ \citenamefont
  {Datta}(2003)}]{response3}%
  \BibitemOpen
  \bibfield  {author} {\bibinfo {author} {\bibfnamefont {M.}~\bibnamefont
  {Paulsson}}\ and\ \bibinfo {author} {\bibfnamefont {S.}~\bibnamefont
  {Datta}},\ }\href {\doibase 10.1103/PhysRevB.67.241403} {\bibfield  {journal}
  {\bibinfo  {journal} {Phys. Rev. B}\ }\textbf {\bibinfo {volume} {67}},\
  \bibinfo {pages} {241403} (\bibinfo {year} {2003})}\BibitemShut {NoStop}%
\bibitem [{\citenamefont {Reddy}\ \emph {et~al.}(2007)\citenamefont {Reddy},
  \citenamefont {Jang}, \citenamefont {Segalman},\ and\ \citenamefont
  {Majumdar}}]{response4}%
  \BibitemOpen
  \bibfield  {author} {\bibinfo {author} {\bibfnamefont {P.}~\bibnamefont
  {Reddy}}, \bibinfo {author} {\bibfnamefont {S.-Y.}\ \bibnamefont {Jang}},
  \bibinfo {author} {\bibfnamefont {R.~A.}\ \bibnamefont {Segalman}}, \ and\
  \bibinfo {author} {\bibfnamefont {A.}~\bibnamefont {Majumdar}},\ }\href
  {\doibase 10.1126/science.1137149} {\bibfield  {journal} {\bibinfo  {journal}
  {Science}\ }\textbf {\bibinfo {volume} {315}},\ \bibinfo {pages} {1568}
  (\bibinfo {year} {2007})}\BibitemShut {NoStop}%
\bibitem [{\citenamefont {Mahan}\ and\ \citenamefont {Sofo}(1996)}]{sofo}%
  \BibitemOpen
  \bibfield  {author} {\bibinfo {author} {\bibfnamefont {G.~D.}\ \bibnamefont
  {Mahan}}\ and\ \bibinfo {author} {\bibfnamefont {J.~O.}\ \bibnamefont
  {Sofo}},\ }\href@noop {} {\bibfield  {journal} {\bibinfo  {journal} {Proc.
  Natl. Acad. Sci. U.S.A.}\ }\textbf {\bibinfo {volume} {93}},\ \bibinfo
  {pages} {7436} (\bibinfo {year} {1996})}\BibitemShut {NoStop}%
\bibitem [{\citenamefont {Snyder}\ \emph {et~al.}(2012)\citenamefont {Snyder},
  \citenamefont {Toberer}, \citenamefont {Khanna},\ and\ \citenamefont
  {Seifert}}]{snyder_thompson}%
  \BibitemOpen
  \bibfield  {author} {\bibinfo {author} {\bibfnamefont {G.~J.}\ \bibnamefont
  {Snyder}}, \bibinfo {author} {\bibfnamefont {E.~S.}\ \bibnamefont {Toberer}},
  \bibinfo {author} {\bibfnamefont {R.}~\bibnamefont {Khanna}}, \ and\ \bibinfo
  {author} {\bibfnamefont {W.}~\bibnamefont {Seifert}},\ }\href {\doibase
  10.1103/PhysRevB.86.045202} {\bibfield  {journal} {\bibinfo  {journal} {Phys.
  Rev. B}\ }\textbf {\bibinfo {volume} {86}},\ \bibinfo {pages} {045202}
  (\bibinfo {year} {2012})}\BibitemShut {NoStop}%
\bibitem [{\citenamefont {Shakouri}\ and\ \citenamefont
  {Bowers}(1997)}]{cooling_ref2}%
  \BibitemOpen
  \bibfield  {author} {\bibinfo {author} {\bibfnamefont {A.}~\bibnamefont
  {Shakouri}}\ and\ \bibinfo {author} {\bibfnamefont {J.~E.}\ \bibnamefont
  {Bowers}},\ }\href {\doibase 10.1063/1.119861} {\bibfield  {journal}
  {\bibinfo  {journal} {Applied Physics Letters}\ }\textbf {\bibinfo {volume}
  {71}},\ \bibinfo {pages} {1234} (\bibinfo {year} {1997})}\BibitemShut
  {NoStop}%
\bibitem [{\citenamefont {Fan}\ \emph {et~al.}(2001)\citenamefont {Fan},
  \citenamefont {Zeng}, \citenamefont {Croke}, \citenamefont {LaBounty},
  \citenamefont {Vashaee}, \citenamefont {Shakouri},\ and\ \citenamefont
  {Bowers}}]{cooling_ref5}%
  \BibitemOpen
  \bibfield  {author} {\bibinfo {author} {\bibfnamefont {X.}~\bibnamefont
  {Fan}}, \bibinfo {author} {\bibfnamefont {G.}~\bibnamefont {Zeng}}, \bibinfo
  {author} {\bibfnamefont {E.}~\bibnamefont {Croke}}, \bibinfo {author}
  {\bibfnamefont {C.}~\bibnamefont {LaBounty}}, \bibinfo {author}
  {\bibfnamefont {D.}~\bibnamefont {Vashaee}}, \bibinfo {author} {\bibfnamefont
  {A.}~\bibnamefont {Shakouri}}, \ and\ \bibinfo {author} {\bibfnamefont
  {J.~E.}\ \bibnamefont {Bowers}},\ }\href {\doibase 10.1049/el:20010096}
  {\bibfield  {journal} {\bibinfo  {journal} {Electronics Letters}\ }\textbf
  {\bibinfo {volume} {37}},\ \bibinfo {pages} {126} (\bibinfo {year}
  {2001})}\BibitemShut {NoStop}%
\bibitem [{\citenamefont {Kim}\ \emph {et~al.}(2010)\citenamefont {Kim},
  \citenamefont {Jeong},\ and\ \citenamefont {Lundstrom}}]{cooling_ref6}%
  \BibitemOpen
  \bibfield  {author} {\bibinfo {author} {\bibfnamefont {R.}~\bibnamefont
  {Kim}}, \bibinfo {author} {\bibfnamefont {C.}~\bibnamefont {Jeong}}, \ and\
  \bibinfo {author} {\bibfnamefont {M.~S.}\ \bibnamefont {Lundstrom}},\ }\href
  {\doibase 10.1063/1.3295899} {\bibfield  {journal} {\bibinfo  {journal}
  {Journal of Applied Physics}\ }\textbf {\bibinfo {volume} {107}},\ \bibinfo
  {pages} {054502} (\bibinfo {year} {2010})}\BibitemShut {NoStop}%
\bibitem [{\citenamefont {Giazotto}\ \emph
  {et~al.}(2006{\natexlab{b}})\citenamefont {Giazotto}, \citenamefont
  {Heikkil\"a}, \citenamefont {Luukanen}, \citenamefont {Savin},\ and\
  \citenamefont {Pekola}}]{cooling_ref9}%
  \BibitemOpen
  \bibfield  {author} {\bibinfo {author} {\bibfnamefont {F.}~\bibnamefont
  {Giazotto}}, \bibinfo {author} {\bibfnamefont {T.~T.}\ \bibnamefont
  {Heikkil\"a}}, \bibinfo {author} {\bibfnamefont {A.}~\bibnamefont
  {Luukanen}}, \bibinfo {author} {\bibfnamefont {A.~M.}\ \bibnamefont {Savin}},
  \ and\ \bibinfo {author} {\bibfnamefont {J.~P.}\ \bibnamefont {Pekola}},\
  }\href@noop {} {\bibfield  {journal} {\bibinfo  {journal} {Rev. Mod. Phys.}\
  }\textbf {\bibinfo {volume} {78}},\ \bibinfo {pages} {217} (\bibinfo {year}
  {2006}{\natexlab{b}})}\BibitemShut {NoStop}%
\bibitem [{\citenamefont {Edwards}\ \emph {et~al.}(1993)\citenamefont
  {Edwards}, \citenamefont {Niu},\ and\ \citenamefont
  {de~Lozanne}}]{cooling_ref10}%
  \BibitemOpen
  \bibfield  {author} {\bibinfo {author} {\bibfnamefont {H.~L.}\ \bibnamefont
  {Edwards}}, \bibinfo {author} {\bibfnamefont {Q.}~\bibnamefont {Niu}}, \ and\
  \bibinfo {author} {\bibfnamefont {A.~L.}\ \bibnamefont {de~Lozanne}},\ }\href
  {\doibase 10.1063/1.110672} {\bibfield  {journal} {\bibinfo  {journal}
  {Applied Physics Letters}\ }\textbf {\bibinfo {volume} {63}},\ \bibinfo
  {pages} {1815} (\bibinfo {year} {1993})}\BibitemShut {NoStop}%
\bibitem [{\citenamefont {Chao}\ \emph {et~al.}(2005)\citenamefont {Chao},
  \citenamefont {Larsson},\ and\ \citenamefont {Mal’shukov}}]{cooling_ref11}%
  \BibitemOpen
  \bibfield  {author} {\bibinfo {author} {\bibfnamefont {K.~A.}\ \bibnamefont
  {Chao}}, \bibinfo {author} {\bibfnamefont {M.}~\bibnamefont {Larsson}}, \
  and\ \bibinfo {author} {\bibfnamefont {A.~G.}\ \bibnamefont {Mal’shukov}},\
  }\href {\doibase 10.1063/1.1992651} {\bibfield  {journal} {\bibinfo
  {journal} {Applied Physics Letters}\ }\textbf {\bibinfo {volume} {87}},\
  \bibinfo {pages} {022103} (\bibinfo {year} {2005})}\BibitemShut {NoStop}%
\bibitem [{\citenamefont {Whitney}(2015)}]{whitney2}%
  \BibitemOpen
  \bibfield  {author} {\bibinfo {author} {\bibfnamefont {R.~S.}\ \bibnamefont
  {Whitney}},\ }\href {\doibase 10.1103/PhysRevB.91.115425} {\bibfield
  {journal} {\bibinfo  {journal} {Phys. Rev. B}\ }\textbf {\bibinfo {volume}
  {91}},\ \bibinfo {pages} {115425} (\bibinfo {year} {2015})}\BibitemShut
  {NoStop}%
\bibitem [{\citenamefont {{Whitney}}(2014)}]{whitney}%
  \BibitemOpen
  \bibfield  {author} {\bibinfo {author} {\bibfnamefont {R.~S.}\ \bibnamefont
  {{Whitney}}},\ }\href {\doibase 10.1103/PhysRevLett.112.130601} {\bibfield
  {journal} {\bibinfo  {journal} {Physical Review Letters}\ }\textbf {\bibinfo
  {volume} {112}},\ \bibinfo {eid} {130601} (\bibinfo {year}
  {2014})}\BibitemShut {NoStop}%
\bibitem [{\citenamefont {Benenti}\ \emph {et~al.}(2017)\citenamefont
  {Benenti}, \citenamefont {Casati}, \citenamefont {Saito},\ and\ \citenamefont
  {Whitney}}]{two}%
  \BibitemOpen
  \bibfield  {author} {\bibinfo {author} {\bibfnamefont {G.}~\bibnamefont
  {Benenti}}, \bibinfo {author} {\bibfnamefont {G.}~\bibnamefont {Casati}},
  \bibinfo {author} {\bibfnamefont {K.}~\bibnamefont {Saito}}, \ and\ \bibinfo
  {author} {\bibfnamefont {R.~S.}\ \bibnamefont {Whitney}},\ }\href {\doibase
  https://doi.org/10.1016/j.physrep.2017.05.008} {\bibfield  {journal}
  {\bibinfo  {journal} {Physics Reports}\ }\textbf {\bibinfo {volume} {694}},\
  \bibinfo {pages} {1 } (\bibinfo {year} {2017})},\ \bibinfo {note}
  {fundamental aspects of steady-state conversion of heat to work at the
  nanoscale}\BibitemShut {NoStop}%
\bibitem [{\citenamefont {Apertet}\ \emph {et~al.}(2013)\citenamefont
  {Apertet}, \citenamefont {Ouerdane}, \citenamefont {Michot}, \citenamefont
  {Goupil},\ and\ \citenamefont {Lecoeur}}]{cooling_ref1}%
  \BibitemOpen
  \bibfield  {author} {\bibinfo {author} {\bibfnamefont {Y.}~\bibnamefont
  {Apertet}}, \bibinfo {author} {\bibfnamefont {H.}~\bibnamefont {Ouerdane}},
  \bibinfo {author} {\bibfnamefont {A.}~\bibnamefont {Michot}}, \bibinfo
  {author} {\bibfnamefont {C.}~\bibnamefont {Goupil}}, \ and\ \bibinfo {author}
  {\bibfnamefont {P.}~\bibnamefont {Lecoeur}},\ }\href@noop {} {\bibfield
  {journal} {\bibinfo  {journal} {EPL (Europhysics Letters)}\ }\textbf
  {\bibinfo {volume} {103}},\ \bibinfo {pages} {40001} (\bibinfo {year}
  {2013})}\BibitemShut {NoStop}%
\bibitem [{\citenamefont {Shakouri}\ \emph {et~al.}(1998)\citenamefont
  {Shakouri}, \citenamefont {Lee}, \citenamefont {Smith}, \citenamefont
  {Narayanamurti},\ and\ \citenamefont {Bowers}}]{cooling_ref3}%
  \BibitemOpen
  \bibfield  {author} {\bibinfo {author} {\bibfnamefont {A.}~\bibnamefont
  {Shakouri}}, \bibinfo {author} {\bibfnamefont {E.~Y.}\ \bibnamefont {Lee}},
  \bibinfo {author} {\bibfnamefont {D.~L.}\ \bibnamefont {Smith}}, \bibinfo
  {author} {\bibfnamefont {V.}~\bibnamefont {Narayanamurti}}, \ and\ \bibinfo
  {author} {\bibfnamefont {J.~E.}\ \bibnamefont {Bowers}},\ }\href {\doibase
  10.1080/108939598200097} {\bibfield  {journal} {\bibinfo  {journal}
  {Microscale Thermophysical Engineering}\ }\textbf {\bibinfo {volume} {2}},\
  \bibinfo {pages} {37} (\bibinfo {year} {1998})}\BibitemShut {NoStop}%
\bibitem [{\citenamefont {Fan}\ \emph {et~al.}(2000)\citenamefont {Fan},
  \citenamefont {Zeng}, \citenamefont {Croke}, \citenamefont {Robinson},
  \citenamefont {LaBounty}, \citenamefont {Shakouri},\ and\ \citenamefont
  {Bowers}}]{cooling_ref4}%
  \BibitemOpen
  \bibfield  {author} {\bibinfo {author} {\bibfnamefont {X.}~\bibnamefont
  {Fan}}, \bibinfo {author} {\bibfnamefont {G.}~\bibnamefont {Zeng}}, \bibinfo
  {author} {\bibfnamefont {E.}~\bibnamefont {Croke}}, \bibinfo {author}
  {\bibfnamefont {G.}~\bibnamefont {Robinson}}, \bibinfo {author}
  {\bibfnamefont {C.}~\bibnamefont {LaBounty}}, \bibinfo {author}
  {\bibfnamefont {A.}~\bibnamefont {Shakouri}}, \ and\ \bibinfo {author}
  {\bibfnamefont {J.~E.}\ \bibnamefont {Bowers}},\ }in\ \href {\doibase
  10.1109/ITHERM.2000.866840} {\emph {\bibinfo {booktitle} {ITHERM 2000. The
  Seventh Intersociety Conference on Thermal and Thermomechanical Phenomena in
  Electronic Systems (Cat. No.00CH37069)}}},\ Vol.~\bibinfo {volume} {1}\
  (\bibinfo {year} {2000})\ p.\ \bibinfo {pages} {307}\BibitemShut {NoStop}%
\bibitem [{\citenamefont {Zhu}\ and\ \citenamefont {Li}(2012)}]{cooling_ref7}%
  \BibitemOpen
  \bibfield  {author} {\bibinfo {author} {\bibfnamefont {J.~p.}\ \bibnamefont
  {Zhu}}\ and\ \bibinfo {author} {\bibfnamefont {G.~x.}\ \bibnamefont {Li}},\
  }\href@noop {} {\bibfield  {journal} {\bibinfo  {journal} {Phys. Rev. A}\
  }\textbf {\bibinfo {volume} {86}},\ \bibinfo {pages} {053828} (\bibinfo
  {year} {2012})}\BibitemShut {NoStop}%
\bibitem [{\citenamefont {Li}\ \emph {et~al.}(2011)\citenamefont {Li},
  \citenamefont {Ouyang}, \citenamefont {Lam},\ and\ \citenamefont
  {You}}]{cooling_ref8}%
  \BibitemOpen
  \bibfield  {author} {\bibinfo {author} {\bibfnamefont {Z.-Z.}\ \bibnamefont
  {Li}}, \bibinfo {author} {\bibfnamefont {S.-H.}\ \bibnamefont {Ouyang}},
  \bibinfo {author} {\bibfnamefont {C.-H.}\ \bibnamefont {Lam}}, \ and\
  \bibinfo {author} {\bibfnamefont {J.~Q.}\ \bibnamefont {You}},\ }\href@noop
  {} {\bibfield  {journal} {\bibinfo  {journal} {EPL (Europhysics Letters)}\
  }\textbf {\bibinfo {volume} {95}},\ \bibinfo {pages} {40003} (\bibinfo {year}
  {2011})}\BibitemShut {NoStop}%
\bibitem [{\citenamefont {Lundstrom}(2002)}]{lundstorm}%
  \BibitemOpen
  \bibfield  {author} {\bibinfo {author} {\bibfnamefont {M.}~\bibnamefont
  {Lundstrom}},\ }\href {http://stacks.iop.org/0957-0233/13/i=2/a=703}
  {\bibfield  {journal} {\bibinfo  {journal} {Measurement Science and
  Technology}\ }\textbf {\bibinfo {volume} {13}},\ \bibinfo {pages} {230}
  (\bibinfo {year} {2002})}\BibitemShut {NoStop}%
\bibitem [{\citenamefont {Datta}(2005)}]{dattabook}%
  \BibitemOpen
  \bibfield  {author} {\bibinfo {author} {\bibfnamefont {S.}~\bibnamefont
  {Datta}},\ }\href@noop {} {\emph {\bibinfo {title} {Quantum Transport:Atom to
  Transistor}}}\ (\bibinfo  {publisher} {Cambridge Press},\ \bibinfo {year}
  {2005})\BibitemShut {NoStop}%
\bibitem [{\citenamefont {Datta}(1997)}]{Datta_Green}%
  \BibitemOpen
  \bibfield  {author} {\bibinfo {author} {\bibfnamefont {S.}~\bibnamefont
  {Datta}},\ }\href@noop {} {\emph {\bibinfo {title} {{Electronic Transport in
  Mesoscopic Systems}}}}\ (\bibinfo  {publisher} {Cambridge University Press},\
  \bibinfo {year} {1997})\BibitemShut {NoStop}%
\bibitem [{\citenamefont {Fetter}\ and\ \citenamefont
  {Walecka}(2003)}]{fetter}%
  \BibitemOpen
  \bibfield  {author} {\bibinfo {author} {\bibfnamefont {A.}~\bibnamefont
  {Fetter}}\ and\ \bibinfo {author} {\bibfnamefont {J.}~\bibnamefont
  {Walecka}},\ }\href {https://books.google.co.in/books?id=0wekf1s83b0C} {\emph
  {\bibinfo {title} {Quantum Theory of Many-particle Systems}}},\ Dover Books
  on Physics\ (\bibinfo  {publisher} {Dover Publications},\ \bibinfo {year}
  {2003})\BibitemShut {NoStop}%
\bibitem [{\citenamefont {Sakurai}\ and\ \citenamefont
  {Napolitano}(2011)}]{sakurai}%
  \BibitemOpen
  \bibfield  {author} {\bibinfo {author} {\bibfnamefont {J.}~\bibnamefont
  {Sakurai}}\ and\ \bibinfo {author} {\bibfnamefont {J.}~\bibnamefont
  {Napolitano}},\ }\href {https://books.google.co.in/books?id=N4I-AQAACAAJ}
  {\emph {\bibinfo {title} {Modern Quantum Mechanics}}}\ (\bibinfo  {publisher}
  {Addison-Wesley},\ \bibinfo {year} {2011})\BibitemShut {NoStop}%
\bibitem [{\citenamefont {Griffiths}(2005)}]{griffith}%
  \BibitemOpen
  \bibfield  {author} {\bibinfo {author} {\bibfnamefont {D.}~\bibnamefont
  {Griffiths}},\ }\href {https://books.google.co.in/books?id=z4fwAAAAMAAJ}
  {\emph {\bibinfo {title} {Introduction to Quantum Mechanics}}},\ Pearson
  international edition\ (\bibinfo  {publisher} {Pearson Prentice Hall},\
  \bibinfo {year} {2005})\BibitemShut {NoStop}%
\bibitem [{\citenamefont {Mingo}\ and\ \citenamefont {Broido}(2004)}]{phonon1}%
  \BibitemOpen
  \bibfield  {author} {\bibinfo {author} {\bibfnamefont {N.}~\bibnamefont
  {Mingo}}\ and\ \bibinfo {author} {\bibfnamefont {D.~A.}\ \bibnamefont
  {Broido}},\ }\href {\doibase 10.1103/PhysRevLett.93.246106} {\bibfield
  {journal} {\bibinfo  {journal} {Phys. Rev. Lett.}\ }\textbf {\bibinfo
  {volume} {93}},\ \bibinfo {pages} {246106} (\bibinfo {year}
  {2004})}\BibitemShut {NoStop}%
\bibitem [{\citenamefont {Mingo}(2004)}]{phonon2}%
  \BibitemOpen
  \bibfield  {author} {\bibinfo {author} {\bibfnamefont {N.}~\bibnamefont
  {Mingo}},\ }\href {\doibase http://dx.doi.org/10.1063/1.1695629} {\bibfield
  {journal} {\bibinfo  {journal} {Applied Physics Letters}\ }\textbf {\bibinfo
  {volume} {84}},\ \bibinfo {pages} {2652} (\bibinfo {year}
  {2004})}\BibitemShut {NoStop}%
\bibitem [{\citenamefont {Zhou}\ \emph {et~al.}(2007)\citenamefont {Zhou},
  \citenamefont {Szczech}, \citenamefont {Pettes}, \citenamefont {Moore},
  \citenamefont {Jin},\ and\ \citenamefont {Shi}}]{phonon3}%
  \BibitemOpen
  \bibfield  {author} {\bibinfo {author} {\bibfnamefont {F.}~\bibnamefont
  {Zhou}}, \bibinfo {author} {\bibfnamefont {J.}~\bibnamefont {Szczech}},
  \bibinfo {author} {\bibfnamefont {M.~T.}\ \bibnamefont {Pettes}}, \bibinfo
  {author} {\bibfnamefont {A.~L.}\ \bibnamefont {Moore}}, \bibinfo {author}
  {\bibfnamefont {S.}~\bibnamefont {Jin}}, \ and\ \bibinfo {author}
  {\bibfnamefont {L.}~\bibnamefont {Shi}},\ }\href {\doibase 10.1021/nl0706143}
  {\bibfield  {journal} {\bibinfo  {journal} {Nano Letters}\ }\textbf {\bibinfo
  {volume} {7}},\ \bibinfo {pages} {1649} (\bibinfo {year} {2007})},\ \bibinfo
  {note} {pMID: 17508772}\BibitemShut {NoStop}%
\bibitem [{\citenamefont {Zhou}\ \emph {et~al.}(2010)\citenamefont {Zhou},
  \citenamefont {Moore}, \citenamefont {Pettes}, \citenamefont {Lee},
  \citenamefont {Seol}, \citenamefont {Ye}, \citenamefont {Rabenberg},\ and\
  \citenamefont {Shi}}]{phonon4}%
  \BibitemOpen
  \bibfield  {author} {\bibinfo {author} {\bibfnamefont {F.}~\bibnamefont
  {Zhou}}, \bibinfo {author} {\bibfnamefont {A.~L.}\ \bibnamefont {Moore}},
  \bibinfo {author} {\bibfnamefont {M.~T.}\ \bibnamefont {Pettes}}, \bibinfo
  {author} {\bibfnamefont {Y.}~\bibnamefont {Lee}}, \bibinfo {author}
  {\bibfnamefont {J.~H.}\ \bibnamefont {Seol}}, \bibinfo {author}
  {\bibfnamefont {Q.~L.}\ \bibnamefont {Ye}}, \bibinfo {author} {\bibfnamefont
  {L.}~\bibnamefont {Rabenberg}}, \ and\ \bibinfo {author} {\bibfnamefont
  {L.}~\bibnamefont {Shi}},\ }\href@noop {} {\bibfield  {journal} {\bibinfo
  {journal} {Journal of Physics D: Applied Physics}\ }\textbf {\bibinfo
  {volume} {43}},\ \bibinfo {pages} {025406} (\bibinfo {year}
  {2010})}\BibitemShut {NoStop}%
\bibitem [{\citenamefont {Boukai}\ \emph {et~al.}(2008)\citenamefont {Boukai},
  \citenamefont {Bunimovich}, \citenamefont {Tahir-Kheli}, \citenamefont {Yu},
  \citenamefont {Goddard},\ and\ \citenamefont {Heath}}]{phonon5}%
  \BibitemOpen
  \bibfield  {author} {\bibinfo {author} {\bibfnamefont {A.~I.}\ \bibnamefont
  {Boukai}}, \bibinfo {author} {\bibfnamefont {Y.}~\bibnamefont {Bunimovich}},
  \bibinfo {author} {\bibfnamefont {J.}~\bibnamefont {Tahir-Kheli}}, \bibinfo
  {author} {\bibfnamefont {J.-K.}\ \bibnamefont {Yu}}, \bibinfo {author}
  {\bibfnamefont {W.~A.}\ \bibnamefont {Goddard}}, \ and\ \bibinfo {author}
  {\bibfnamefont {J.~R.}\ \bibnamefont {Heath}},\ }\href {\doibase
  10.1038/nature06458} {\bibfield  {journal} {\bibinfo  {journal} {Nature}\
  }\textbf {\bibinfo {volume} {451}},\ \bibinfo {pages} {168} (\bibinfo {year}
  {2008})}\BibitemShut {NoStop}%
\bibitem [{\citenamefont {Hochbaum}\ \emph {et~al.}(2008)\citenamefont
  {Hochbaum}, \citenamefont {Chen}, \citenamefont {Delgado}, \citenamefont
  {Liang}, \citenamefont {Garnett}, \citenamefont {Najarian}, \citenamefont
  {Majumdar},\ and\ \citenamefont {Yang}}]{phonon6}%
  \BibitemOpen
  \bibfield  {author} {\bibinfo {author} {\bibfnamefont {A.~I.}\ \bibnamefont
  {Hochbaum}}, \bibinfo {author} {\bibfnamefont {R.}~\bibnamefont {Chen}},
  \bibinfo {author} {\bibfnamefont {R.~D.}\ \bibnamefont {Delgado}}, \bibinfo
  {author} {\bibfnamefont {W.}~\bibnamefont {Liang}}, \bibinfo {author}
  {\bibfnamefont {E.~C.}\ \bibnamefont {Garnett}}, \bibinfo {author}
  {\bibfnamefont {M.}~\bibnamefont {Najarian}}, \bibinfo {author}
  {\bibfnamefont {A.}~\bibnamefont {Majumdar}}, \ and\ \bibinfo {author}
  {\bibfnamefont {P.}~\bibnamefont {Yang}},\ }\href {\doibase
  10.1038/nature06381} {\bibfield  {journal} {\bibinfo  {journal} {Nature
  Publishing Group}\ }\textbf {\bibinfo {volume} {451}},\ \bibinfo {pages}
  {163} (\bibinfo {year} {2008})}\BibitemShut {NoStop}%
\bibitem [{\citenamefont {Balandin}\ \emph {et~al.}(1999)\citenamefont
  {Balandin}, \citenamefont {Khitun}, \citenamefont {Liu}, \citenamefont
  {Wang}, \citenamefont {Borca-Tasciuc},\ and\ \citenamefont {Chen}}]{phonon7}%
  \BibitemOpen
  \bibfield  {author} {\bibinfo {author} {\bibfnamefont {A.}~\bibnamefont
  {Balandin}}, \bibinfo {author} {\bibfnamefont {A.}~\bibnamefont {Khitun}},
  \bibinfo {author} {\bibfnamefont {J.}~\bibnamefont {Liu}}, \bibinfo {author}
  {\bibfnamefont {K.}~\bibnamefont {Wang}}, \bibinfo {author} {\bibfnamefont
  {T.}~\bibnamefont {Borca-Tasciuc}}, \ and\ \bibinfo {author} {\bibfnamefont
  {G.}~\bibnamefont {Chen}},\ }in\ \href {\doibase 10.1109/ICT.1999.843365}
  {\emph {\bibinfo {booktitle} {Eighteenth International Conference on
  Thermoelectrics}}}\ (\bibinfo {year} {1999})\ pp.\ \bibinfo {pages}
  {189--192}\BibitemShut {NoStop}%
\bibitem [{\citenamefont {Chen}(1998)}]{superlattice1}%
  \BibitemOpen
  \bibfield  {author} {\bibinfo {author} {\bibfnamefont {G.}~\bibnamefont
  {Chen}},\ }\href {\doibase 10.1103/PhysRevB.57.14958} {\bibfield  {journal}
  {\bibinfo  {journal} {Phys. Rev. B}\ }\textbf {\bibinfo {volume} {57}},\
  \bibinfo {pages} {14958} (\bibinfo {year} {1998})}\BibitemShut {NoStop}%
\bibitem [{\citenamefont {Koga}\ \emph {et~al.}(2000)\citenamefont {Koga},
  \citenamefont {Cronin}, \citenamefont {Dresselhaus}, \citenamefont {Liu},\
  and\ \citenamefont {Wang}}]{superlattice2}%
  \BibitemOpen
  \bibfield  {author} {\bibinfo {author} {\bibfnamefont {T.}~\bibnamefont
  {Koga}}, \bibinfo {author} {\bibfnamefont {S.~B.}\ \bibnamefont {Cronin}},
  \bibinfo {author} {\bibfnamefont {M.~S.}\ \bibnamefont {Dresselhaus}},
  \bibinfo {author} {\bibfnamefont {J.~L.}\ \bibnamefont {Liu}}, \ and\
  \bibinfo {author} {\bibfnamefont {K.~L.}\ \bibnamefont {Wang}},\ }\href@noop
  {} {\bibfield  {journal} {\bibinfo  {journal} {Applied Physics Letters}\
  }\textbf {\bibinfo {volume} {77}} (\bibinfo {year} {2000})}\BibitemShut
  {NoStop}%
\bibitem [{\citenamefont {Davis}\ and\ \citenamefont
  {Hussein}(2014)}]{nanoflake_heat}%
  \BibitemOpen
  \bibfield  {author} {\bibinfo {author} {\bibfnamefont {B.~L.}\ \bibnamefont
  {Davis}}\ and\ \bibinfo {author} {\bibfnamefont {M.~I.}\ \bibnamefont
  {Hussein}},\ }\href {\doibase 10.1103/PhysRevLett.112.055505} {\bibfield
  {journal} {\bibinfo  {journal} {Phys. Rev. Lett.}\ }\textbf {\bibinfo
  {volume} {112}},\ \bibinfo {pages} {055505} (\bibinfo {year}
  {2014})}\BibitemShut {NoStop}%
\bibitem [{\citenamefont {Pan}\ \emph {et~al.}(2015)\citenamefont {Pan},
  \citenamefont {Hong}, \citenamefont {Raja}, \citenamefont {Zimmermann},
  \citenamefont {Tiwari},\ and\ \citenamefont {Poulikakos}}]{nanowire_heat1}%
  \BibitemOpen
  \bibfield  {author} {\bibinfo {author} {\bibfnamefont {Y.}~\bibnamefont
  {Pan}}, \bibinfo {author} {\bibfnamefont {G.}~\bibnamefont {Hong}}, \bibinfo
  {author} {\bibfnamefont {S.~N.}\ \bibnamefont {Raja}}, \bibinfo {author}
  {\bibfnamefont {S.}~\bibnamefont {Zimmermann}}, \bibinfo {author}
  {\bibfnamefont {M.~K.}\ \bibnamefont {Tiwari}}, \ and\ \bibinfo {author}
  {\bibfnamefont {D.}~\bibnamefont {Poulikakos}},\ }\href {\doibase
  10.1063/1.4913879} {\bibfield  {journal} {\bibinfo  {journal} {Applied
  Physics Letters}\ }\textbf {\bibinfo {volume} {106}},\ \bibinfo {pages}
  {093102} (\bibinfo {year} {2015})}\BibitemShut {NoStop}%
\bibitem [{\citenamefont {Feser}\ \emph {et~al.}(2012)\citenamefont {Feser},
  \citenamefont {Sadhu}, \citenamefont {Azeredo}, \citenamefont {Hsu},
  \citenamefont {Ma}, \citenamefont {Kim}, \citenamefont {Seong}, \citenamefont
  {Fang}, \citenamefont {Li}, \citenamefont {Ferreira}, \citenamefont {Sinha},\
  and\ \citenamefont {Cahill}}]{nanowire_heat2}%
  \BibitemOpen
  \bibfield  {author} {\bibinfo {author} {\bibfnamefont {J.~P.}\ \bibnamefont
  {Feser}}, \bibinfo {author} {\bibfnamefont {J.~S.}\ \bibnamefont {Sadhu}},
  \bibinfo {author} {\bibfnamefont {B.~P.}\ \bibnamefont {Azeredo}}, \bibinfo
  {author} {\bibfnamefont {K.~H.}\ \bibnamefont {Hsu}}, \bibinfo {author}
  {\bibfnamefont {J.}~\bibnamefont {Ma}}, \bibinfo {author} {\bibfnamefont
  {J.}~\bibnamefont {Kim}}, \bibinfo {author} {\bibfnamefont {M.}~\bibnamefont
  {Seong}}, \bibinfo {author} {\bibfnamefont {N.~X.}\ \bibnamefont {Fang}},
  \bibinfo {author} {\bibfnamefont {X.}~\bibnamefont {Li}}, \bibinfo {author}
  {\bibfnamefont {P.~M.}\ \bibnamefont {Ferreira}}, \bibinfo {author}
  {\bibfnamefont {S.}~\bibnamefont {Sinha}}, \ and\ \bibinfo {author}
  {\bibfnamefont {D.~G.}\ \bibnamefont {Cahill}},\ }\href {\doibase
  10.1063/1.4767456} {\bibfield  {journal} {\bibinfo  {journal} {Journal of
  Applied Physics}\ }\textbf {\bibinfo {volume} {112}},\ \bibinfo {pages}
  {114306} (\bibinfo {year} {2012})}\BibitemShut {NoStop}%
\end{thebibliography}%


\providecommand{\noopsort}[1]{}\providecommand{\singleletter}[1]{#1}%
%

\end{document}